\documentstyle[aps,epsf,twocolumn]{revtex}

\title{\bf Correlation property of length sequences  based on global
structure of complete genome}
\author{Zu-Guo Yu$^{1,2}$\thanks{Corresponding author, e-mail: yuzg@hotmail.com or
 z.yu@qut.edu.au}, V. V. Anh$^{1}$ and Bin Wang$^{3}$\\
 {\small $^1$Centre in Statistical Science and Industrial Mathematics, Queensland University} \\
 {\small of Technology, GPO Box 2434, Brisbane, Q 4001, Australia.}\\
 {\small $^2$Department of Mathematics, Xiangtan University, Hunan 411105, P. R. China.\thanks{
 This is the permanent corresponding address of Zu-Guo Yu.}}\\
 {\small $^{3}$Institute of Theoretical Physics,  Academia Sinica,
  P.O. Box 2735, Beijing 100080, P. R. China.}
 }
\newcommand{\be}{\begin{equation}}
\newcommand{\ee}{\end{equation}}
\begin{document}
\maketitle

\begin{abstract}
{\bf Abstract--} This paper considers three kinds of length sequences of the complete genome.
Detrended fluctuation analysis, spectral analysis, and the mean distance
spanned within time $L$ are used to discuss the correlation property of
these sequences.  The
values of the exponents from these methods of these three kinds of length sequences of bacteria
indicate that the long-range correlations exist in most of these sequences. 
The correlation have a rich variety of behaviours including
the presence of anti-correlations. Further more, using the exponent $\gamma$, it is found that
these correlations are all linear ($\gamma=1.0\pm 0.03$). It is also found 
that these sequences exhibit $1/f$ noise in
some interval of frequency ($f>1$). The length of this interval of frequency
depends on the length of the sequence. The shape of the periodogram in $f>1$
exhibits some periodicity. The period seems to depend on the length and the
complexity of the length sequence.
\end{abstract}

\section{Introduction}

\ \ Recently, there has been considerable interest in the finding of
long-range correlation (LRC) in DNA sequences $[1-16]$. Li {\it et al}$^{\cite
{li}}$ found that the spectral density of a DNA sequence containing mostly
introns shows $1/f^{\beta }$ behaviour, which indicates the presence of LRC.
The correlation properties of coding and noncoding DNA sequences were first
studied by Peng {\it et al}$^{\cite{peng}}$ in their fractal landscape or
DNA walk model. The DNA walk defined in \cite{peng} is that the walker steps
``up'' if a pyrimidine ($C$ or $T$) occurs at position $i$ along the DNA
chain, while the walker steps ``down'' if a purine ($A$ or $G$) occurs at
position $i$. Peng {\it et al}$^{\cite{peng}}$ discovered that there exists
LRC in noncoding DNA sequences while the coding sequences correspond to a
regular random walk. By doing a more detailed analysis, Chatzidimitriou,
Dreismann and Larhammar$^{\cite{CDL93}}$ concluded that both coding and
noncoding sequences exhibit LRC. A subsequent work by Prabhu and Claverie$^{
\cite{PC92}}$ also substantially corroborates these results. If one
considers more details by distinguishing $C$ from $T$ in pyrimidine, and $A$
from $G$ in purine (such as two or three-dimensional DNA walk model$^{\cite
{luo}}$ and maps given in \cite{YC}), then the presence of base correlation
has been found even in coding sequences. In view of the controversy about
the presence of correlation in all DNA or only in noncoding DNA, Buldyrev
{\it et al}$^{\cite{Bul95}}$ showed the LRC appears mainly in noncoding
DNA using all the DNA sequences available.
Alternatively, Voss$^{\cite{voss}}$,
based on equal-symbol correlation, showed a power law behaviour for the
sequences studied regardless of the percent of intron contents.
Investigations based on different models seem to suggest different results,
as they all look into only a certain aspect of the entire DNA sequence. It
is therefore important to investigate the degree of correlations in a
model-independent way.

Since the first complete genome of the free-living bacterium {\it Mycoplasma
genitalium} was sequenced in 1995$^{\cite{Fraser}}$, an ever-growing number
of complete genomes has been deposited in public databases. The availability
of complete genomes induces the possibility to ask some global questions on
these sequences. The avoided and under-represented strings in some bacterial
complete genomes have been discussed in \cite{yhxc99,hlz98,hxyc99}. A time
series model of CDS in complete genome has also been proposed in \cite{YW99}.
Maria de Sousa Vieira$^{\cite{Vie99}}$ have done the low-frequency analysis 
of complete
DNA of 13 microbial genomes and shown that its fractal behaviour not always
prevails through the entire chain, the autocorrelation function have a rich
variety of behaviours including the presence of anti-correlations.

For the importance of the numbers, sizes and ordering of genes along the
chromosome, one can refer to Part 5 of the famous book of Lewin (Ref.\cite{Lew97}).
Hence one may ignore the composition of the four kinds of bases in coding and
noncoding segments and only considers the rough structure of the complete
genome or long DNA sequences. Provata and Almirantis $^{\cite{PY}}$ proposed
a fractal Cantor pattern of DNA. They map coding segments to filled regions
and noncoding segments to empty regions of random Cantor set and then
calculate the fractal dimension of the random fractal set. They found that
the coding/noncoding partition in DNA sequences of lower organisms is
homogeneous-like, while in the higher eucariotes the partition is fractal.
This result seems too rough to distinguish bacteria because the fractal
dimensions of bacteria they gave out are all the same. The classification
and evolution relationship of bacteria is one of the most important problem
in DNA research. Yu and Anh$^{\cite{YA00}}$ proposed a time series model
based on the global structure of the complete genome and considered three
kinds of length sequences. After calculating the correlation dimensions and
Hurst exponents, it was found that one can get more information from this
model than that of fractal Cantor pattern. Some results on the
classification and evolution relationship of bacteria were found in \cite
{YA00}. Naturally it is desirable to know if there exists LRC in these
length sequences. 
The quantification of these correlations could give an 
insight of the role of the ordering of genes on the chromosome, which
is far to be irrelevant for gene function. 
We attempt to answer this question in this paper.

Viewing from the level of structure, the complete genome of an organism is
made up of coding and noncoding segments. Here the length of a
coding/noncoding segment means the number of its bases. Based on the lengths
of coding/noncoding segments in the complete genome, we can get three kinds
of integer sequences by the following ways.

i) First we order all lengths of coding and noncoding segments according to
the order of coding and noncoding segments in the complete genome, then
replace the lengths of noncoding segments by their negative numbers. This
allows to distinguish lengths of coding and noncoding segments. This integer
sequence is named {\it whole length sequence}.

ii) We order all lengths of coding segments according to the order of coding
segments in the complete genome. We name this integer sequence {\it coding
length sequence}.

iii) We order all lengths of noncoding segments according to the order of
noncoding segments in the complete genome. This integer sequence is named 
{\it noncoding length sequence}.

We can now view these three kinds of integer sequences as time series. In
the following, we will investigate the correlation property through {\it %
Detrended Fluctuation Analysis} (DFA)$^{\cite{GPH}}$ and spectral analysis.

\begin{figure}
\centerline{\epsfxsize=8cm \epsfbox{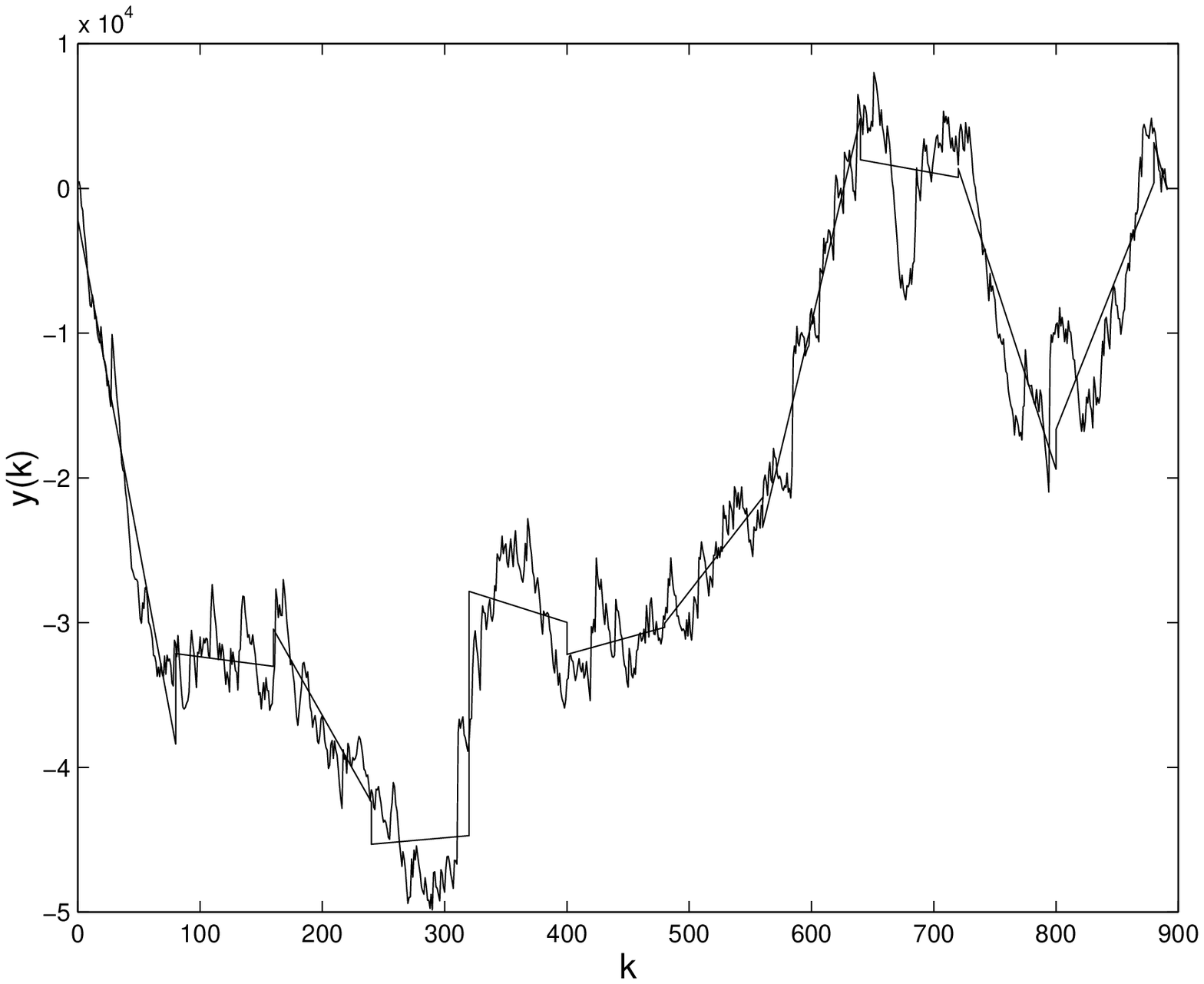}
 \epsfxsize=8cm \epsfbox{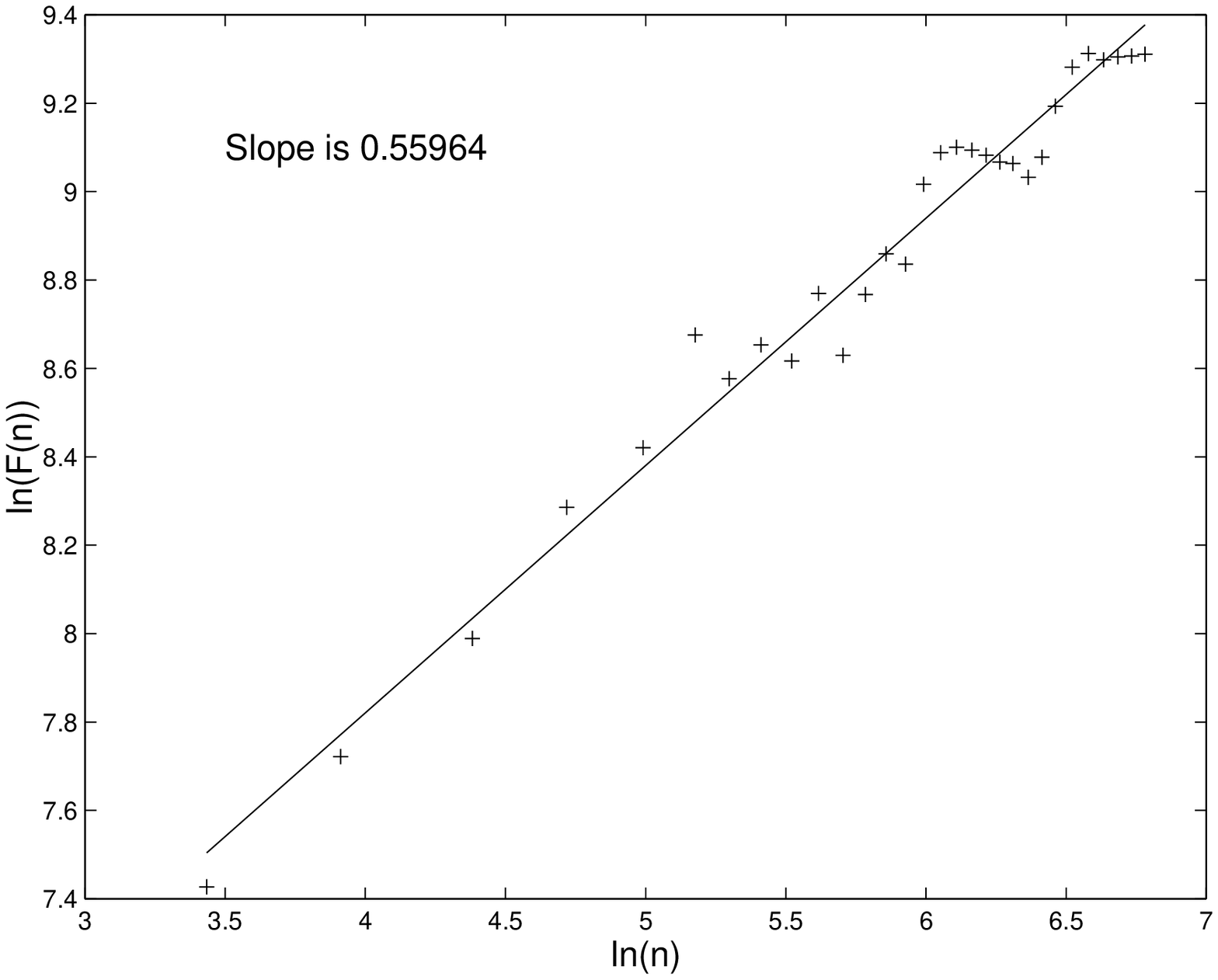}} 
\caption{\footnotesize An example to show how to do detrended fluctuation analysis, {\bf Left)}
To get the sequence $y_n(k)$; {\bf Right)} To get the exponent $\alpha$ using
least-square linear fit. }
\label{jffig}
\end{figure}

\section{Detrended fluctuation analysis and spectral analysis}

\ \ \ \ We denote a time series as $X(t),t=1,\cdots ,N$. First the time
series is integrated as $y(k)=\sum_{t=1}^{k}[X(t)-X_{ave}]$, where $X_{ave}$
is the average over the whole time period. Next, the integrated time series
is divided into boxes of equal length, $n$. In each box of length $n$, a
least-squares line is fit to the data, representing the trend in that box.
The $y$ coordinate of the straight line segments is denoted by $y_{n}(k)$.
We then detrend the integrated time series, $y(k)$, by subtracting the local
trend, $y_{n}(k)$, in each box. The root-mean-square fluctuation of this
integrated and detrended time series is calculated as 
\begin{equation}
F(n)=\sqrt{\frac{1}{N}\sum_{k=1}^{N}[y(k)-y_{n}(k)]^{2}}
\end{equation}
Typically, $F(n)$ will increase with box size $n$. A linear relationship on
a double log graph indicates the presence of scaling 
\begin{equation}
F(n)\quad \propto \quad n^{\alpha }.
\end{equation}
Under such conditions, the fluctuations can be characterised by the scaling
exponent $\alpha $, the slope of the line relating $\ln F(n)$ to $\ln \ n$.
For uncorrelated data, the integrated value, $y(k)$ corresponds to a random
walk, and therefore, $\alpha =0.5$. An value of $0.5<\alpha <
1.0$ indicates the presence of LRC so that a large interval is more likely
to be followed by a large interval and vice versa. In contrast, $0<\alpha
<0.5$ indicates a different type of power-law correlations such that large
and small values of time series are more likely to alternate. For examples,
we give the DFA of the coding length sequence of {\it A. aeolicus} in
FIG. \ref{jffig}.

\begin{figure}
\centerline{\epsfxsize=8cm \epsfbox{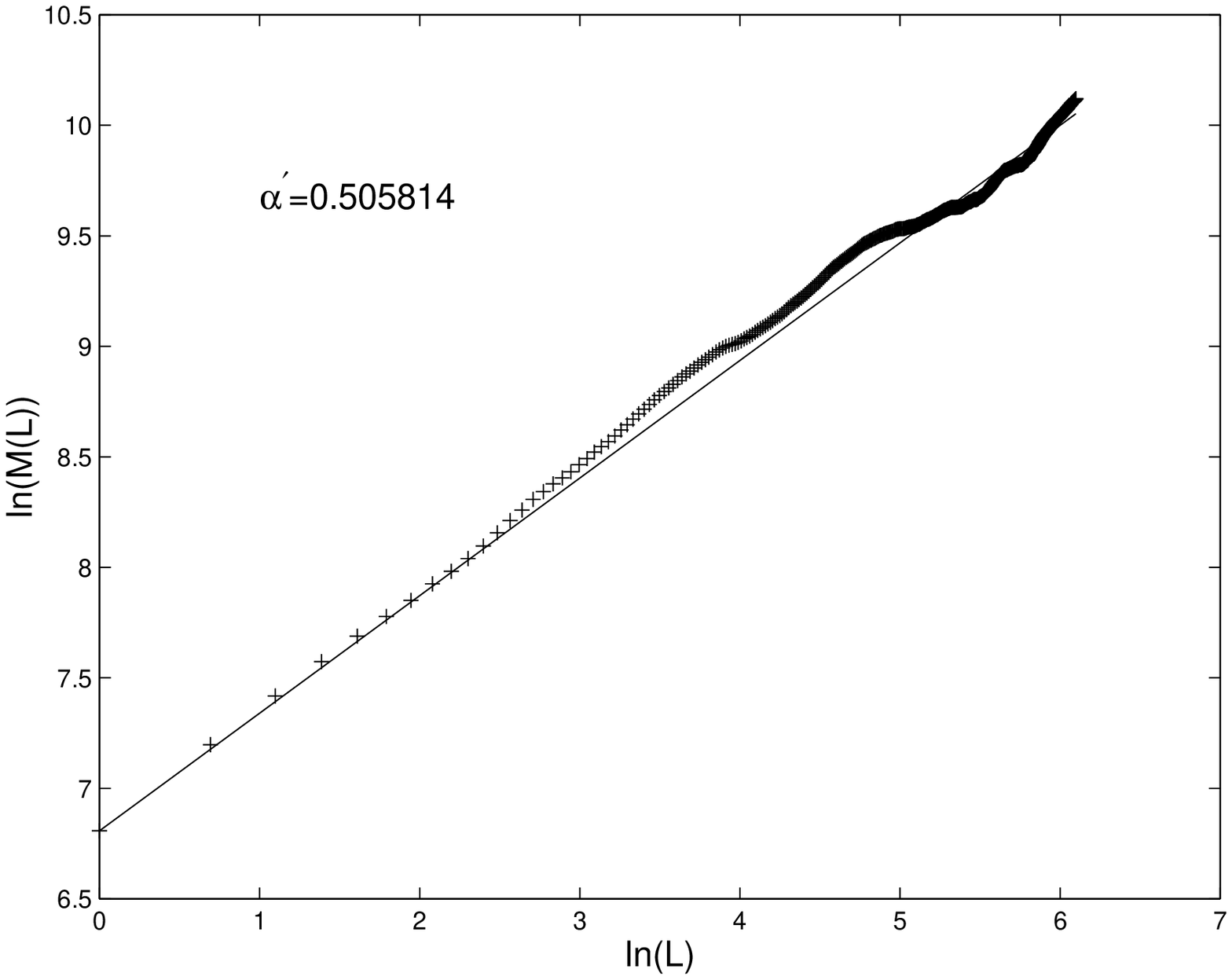}
 \epsfxsize=8cm \epsfbox{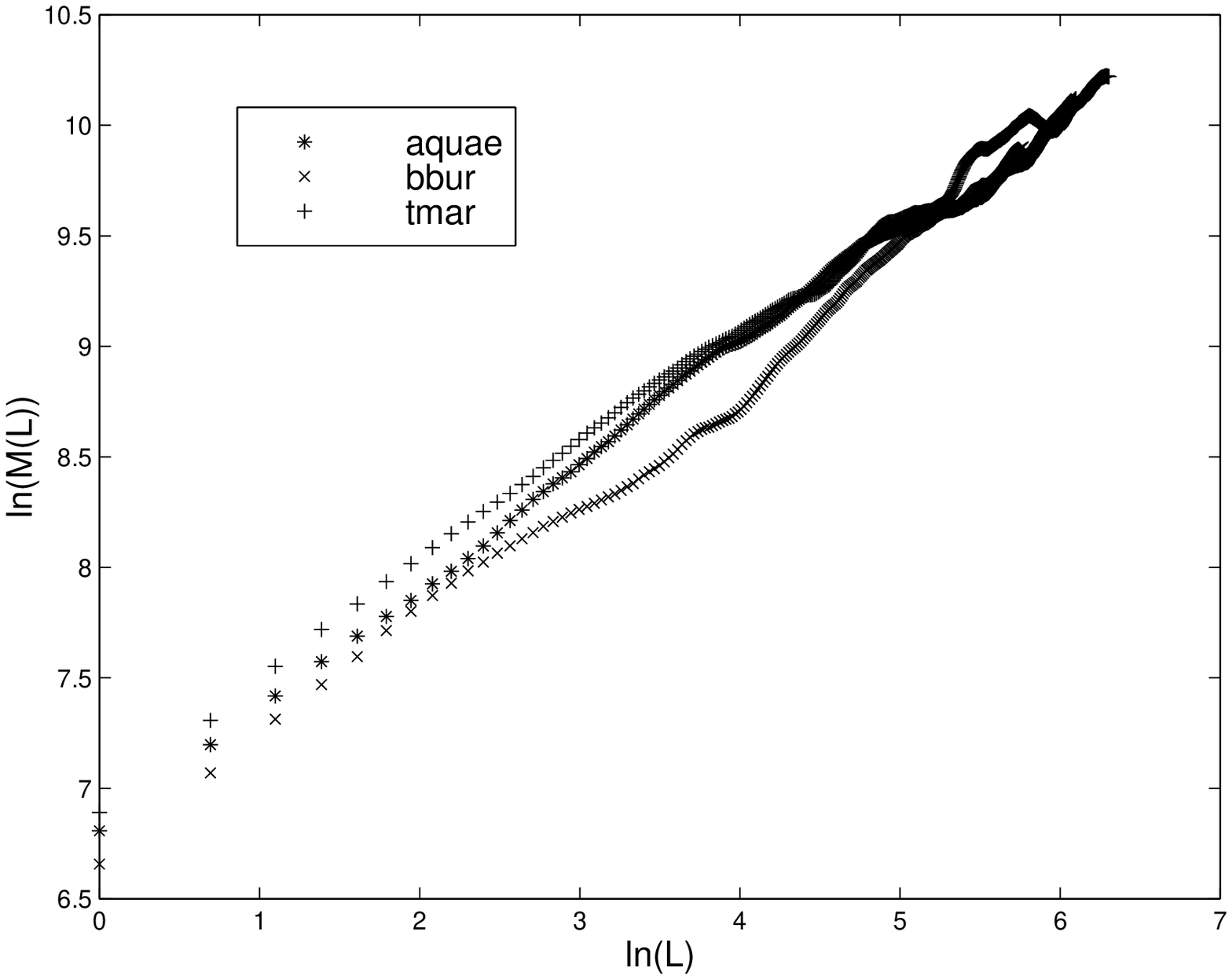}} 
\caption{\footnotesize  {\bf Left)} To get the exponent $\alpha^{\prime}$ using
least-square linear fit; {\bf Right)}  The analysis of  coding length sequences
of three bacteria using mean distance a walker spanned within time $L$.}
\label{MLfig}
\end{figure}

 Now we analyse the time series using the quantity $M(L)$, the mean distance
a walker spanned within time $L$. Dunki and Ambuhl$^{\cite{Dun1,Dun2}}$ used
this quantity to discuss the scaling property in temporal patterns of
schizophrenia. Denote 
\begin{equation}
W(j):=\sum_{t=1}^{j}[X(t)-X_{ave}],
\end{equation}
from which we get the walks 
\begin{equation}
M(L):=<|W(j)-W(j+L)|>_{j},  \label{M3}
\end{equation}
where $<\,>_{j}$ denotes the average over $j$, and $j=1,\cdots ,N-L$. The
time shift $L$ typically varies from $1,\cdots ,N/2$. From a physics
viewpoint, $M(L)$ might be thought of as the variance evolution of a random
walker's total displacement mapped from the time series $X(t)$. $M(L)$ may
be assessed for LRC $^{\cite{peng2}}$ (e.g. $M(L)\ \propto \ L^{\alpha
^{\prime }}$, $\alpha ^{\prime }=1/2$ corresponding to the random case). We
give some examples to estimate the scale parameter $\alpha ^{\prime }$ in
 FIG. \ref{MLfig}.

Dunki {\it et al}$^{\cite{Dun2}}$
proposed the following scale which seems to perform better than the scale $%
\alpha ^{\prime }$. The definition

\begin{equation}
W^{\prime}(j):=\sum_{t=1}^j|X(t)-X_{ave}|
\end{equation}
leads to

\begin{equation}
M^{\prime }(L):=<|W^{\prime }(j)-W^{\prime }(j+L)|>_{j}.  \label{M4}
\end{equation}
Analyses of test time series showed that (\ref{M4}) are more robust against
distortion or discretization of the corresponding amplitudes $X(t)$ than (%
\ref{M3}). From the $\ln (L)$ v.s. $\ln (M^{\prime }(L))$ plane, we find the
relation 
\begin{equation}
M^{\prime }(L)\quad \propto \quad L^{\gamma }.
\end{equation}

\begin{figure}
\centerline{\epsfxsize=8cm \epsfbox{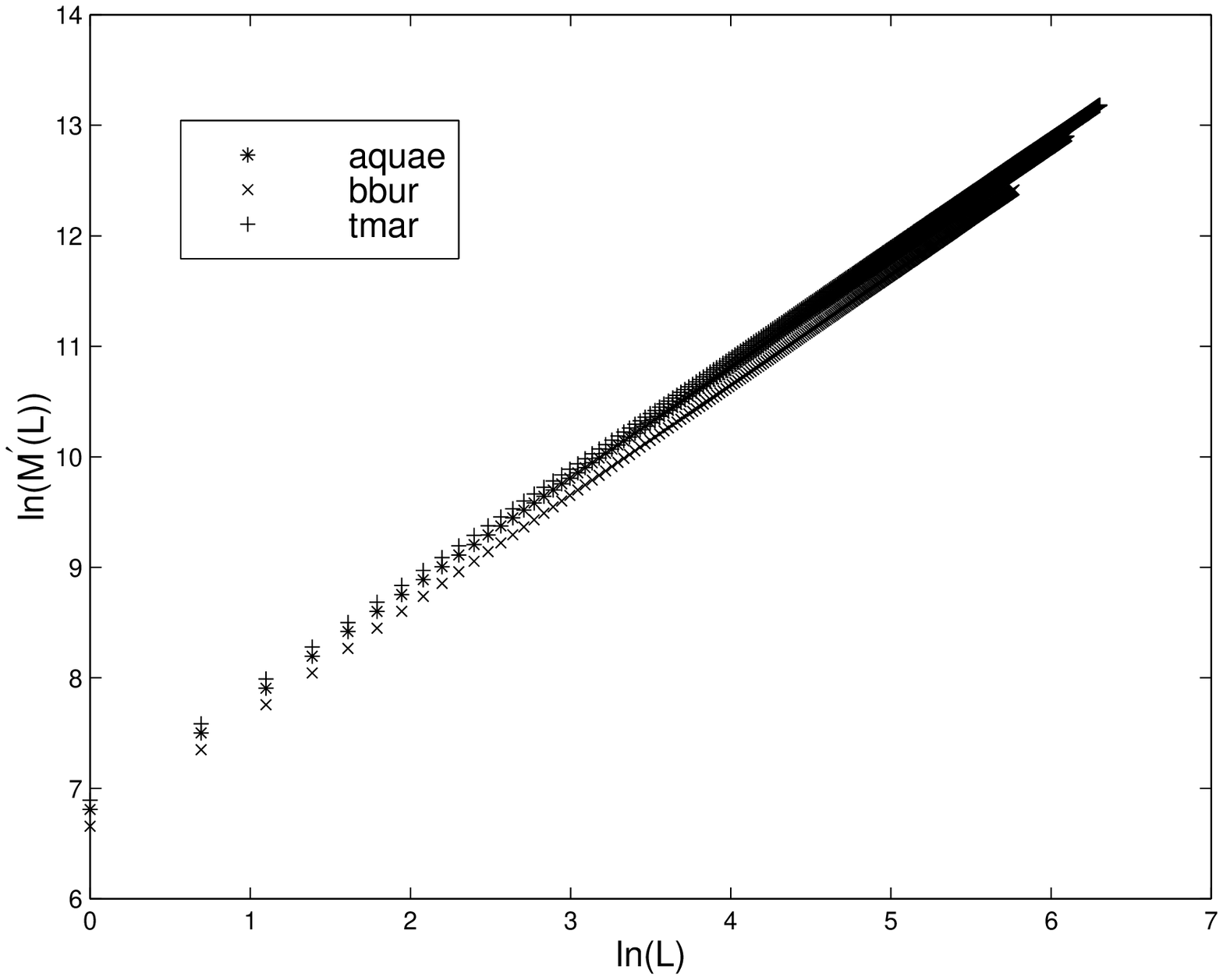}
  \epsfxsize=8cm \epsfbox{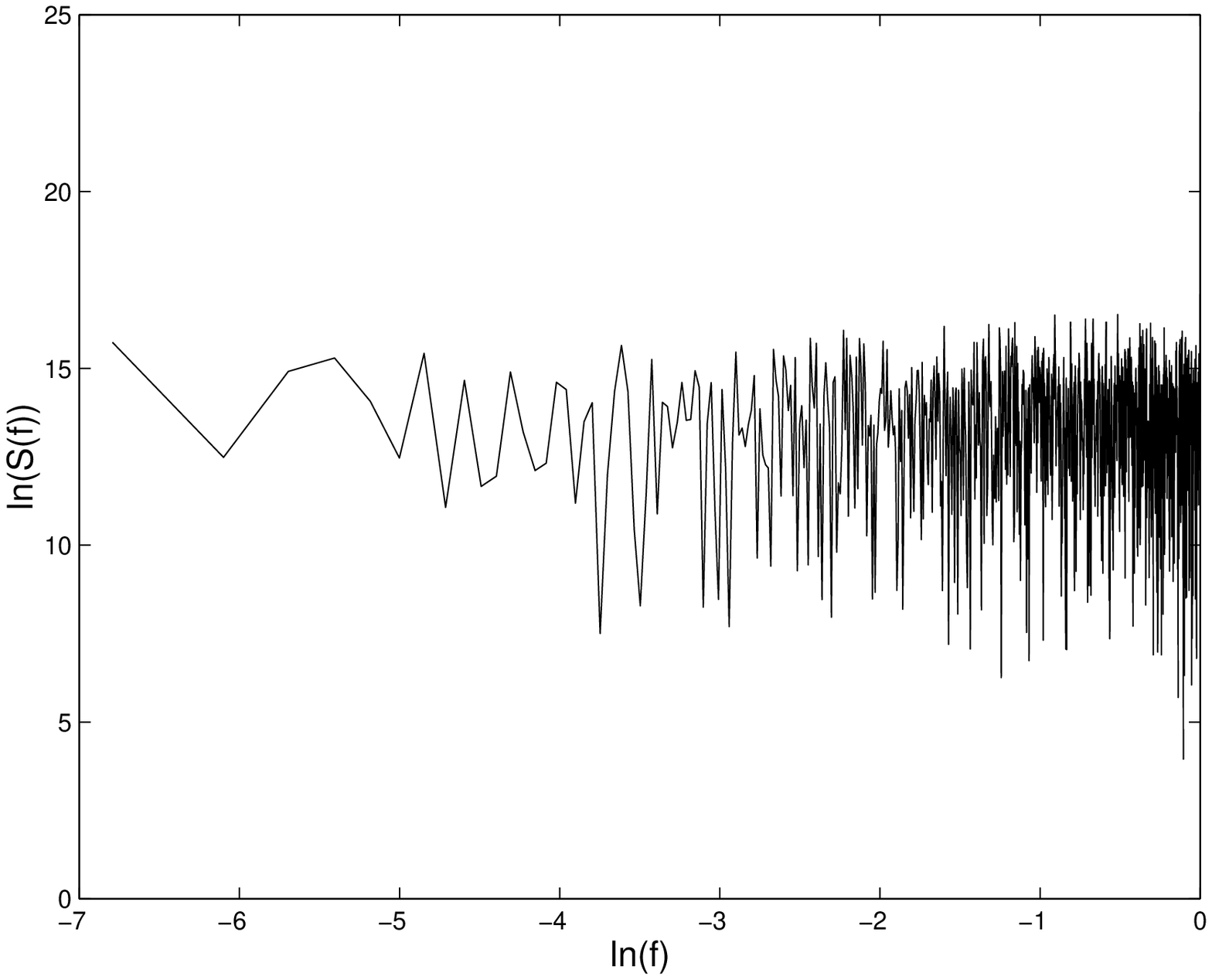}
  }
\caption{\footnotesize  
{\bf Left}) Estimate the scale $\protect\gamma$. {\bf Right}) An example of 
spectral analysis of low frequencies $f<1$.}
\label{M1pufig}
\end{figure}

The exponent $\gamma$ measures only the presence of nonlinear correlations
and remains equal to unity for all sequences with only linear correlations.
 We  carried out this kind of analysis on coding
length sequences of {\it A. aeolicus, B. burgdorferi} and {\it T. maritima}. 
The results
are reported in the left figure of FIG. \ref{M1pufig}.

 We also consider the discrete Fourier transform$^{\cite{Shu88}}$ of the time
series $X(t),t=1,\cdots ,N$ defined by 
\begin{equation}
\widehat{X}(f)=N^{-\frac{1}{2}}\sum_{t=0}^{N-1}X(t+1)e^{-2\pi ift},
\end{equation}
then 
\begin{equation}
S(f)=|\widehat{X}(f)|^{2}
\end{equation}
is the {\it power spectrum of }$X\left( t\right) $. In recent studies, it
has been found $^{\cite{Rob74}}$ that many natural phenomena lead to the
power spectrum of the form $1/f^{\beta }$. This kind of dependence was named 
$1/f$ noise, in contrast to white noise $S(f)=const$, i.e. $\beta =0$. Let
the frequency $f$ take $k$ values $f_{k}=k/N,k=1,\cdots ,N$. From the $\ln
(f)$ v.s. $\ln (S(f))$ graph, the existence of $1/f^{\beta }\ $doesn't seem
apparent. For example, we give the figure of the coding length sequence of
{\it A. aeolicus} on the right of FIG. \ref{M1pufig}.

When we use the least squares line to fit data, we need to consider the
errors. If the data are $\{(x_i,y_i)\}_{i=1}^n$, we can define the {\it
coefficient of linear correlation} as$^{\cite{RS85}}$
\be
r=\frac{\sum_{i=1}^n (x_i-\bar{x})(y_i-\bar{y})}
{\sqrt(\sum_{i=1}^n (x_i-\bar{x})^2\sum_{i=1}^n (y_i-\bar{y})^2)},
\ee
where $\bar{x}$ and $\bar{y}$ are the average of the values $\{x_i\}_{i=1}^n$
and $\{y_i\}_{i=1}^n$ respectively. If $r=\pm 1$, then the points lie exactly on a straight line; that is, there is
a perfect linear relationship between $x$ and $y$. If $r=0$, there is no linear 
relationship. The quantity $r$ measures the strength of linear relationships between
$x$ and $y$. The values of $r$ in figures of obtaining exponents $\alpha$,
$\alpha^{\prime}$ and $\beta$ are 0.987685, 0.9949939 and 3.7918E-03 respectively.
Hence we can see the informations given by exponents $\alpha$ and
$\alpha^{\prime}$ are more convincing than that given by exponent $\beta$.

\section{ Data and results.}

\ \ \ \ More than 21 bacterial complete genomes are now available in public
databases. There are five Archaebacteria: {\it Archaeoglobus fulgidus} (aful),
{\it Pyrococcus abyssi} (pabyssi), {\it Methanococcus jannaschii} (mjan), 
{\it Aeropyrum
pernix} (aero) and {\it Methanobacterium thermoautotrophicum} (mthe); four
Gram-positive Eubacteria: {\it Mycobacterium tuberculosis} (mtub), {\it Mycoplasma
pneumoniae} (mpneu), {\it Mycoplasma genitalium} (mgen), and {\it Bacillus 
subtilis}
(bsub). The others are Gram-negative Eubacteria. These consist of two
Hyperthermophilic bacteria: {\it Aquifex aeolicus} (aquae) and {\it Thermotoga 
maritima}
(tmar); six Proteobacteria: {\it Rhizobium sp. NGR234} (pNGR234), {\it 
Escherichia coli}
(ecoli), {\it Haemophilus influenzae} (hinf), {\it Helicobacter pylori J99} (hpyl99),
{\it Helicobacter pylori 26695} (hpyl) and {\it Rickettsia prowazekii} (rpxx); 
two
Chlamydia: {\it Chlamydia trachomatis} (ctra) and {\it Chlamydia pneumoniae} (cpneu), and
two Spirochaete: {\it Borrelia burgdorferi} (bbur) and {\it Treponema pallidum} (tpal).

We calculate scales $\alpha $, $\alpha^{\prime}$, $\beta $ of low 
frequencies ($f<1$) and $%
\gamma $ of three kinds of length sequences of the above 21 bacteria. The
estimated results are given in Table \ref{jf} ( where we denote by $\alpha
_{whole}$, $\alpha _{cod}$ and $\alpha _{noncod}$ the scales of DFA of the
whole, coding and noncoding length sequences, from top to bottom, in the
increasing order of the value of $\alpha _{whole}$ ), Table \ref{ML} ( where 
we denote by $\alpha^{\prime}
_{whole}$, $\alpha^{\prime} _{cod}$ and $\alpha^{\prime} _{noncod}$ 
the scales of $M(L)$
 of the
whole, coding and noncoding length sequences, from top to bottom, in the
increasing order of the value of $\alpha^{\prime} _{whole}$ ) and 
Table \ref{puM} (
where we denote by $\beta _{whole}$, $\beta _{cod}$ and $\beta _{noncod}$
the scales of spectral analysis of the whole, coding and noncoding length
sequences, from top to bottom, in the decreasing order of the value of $%
\beta _{whole}$; we denote by $\gamma _{whole}$, $\gamma _{cod}$ and $\gamma
_{noncod}$ the scales of $\gamma $ of the whole, coding and noncoding length
sequences).

\begin{figure}
\centerline{\epsfxsize=8cm \epsfbox{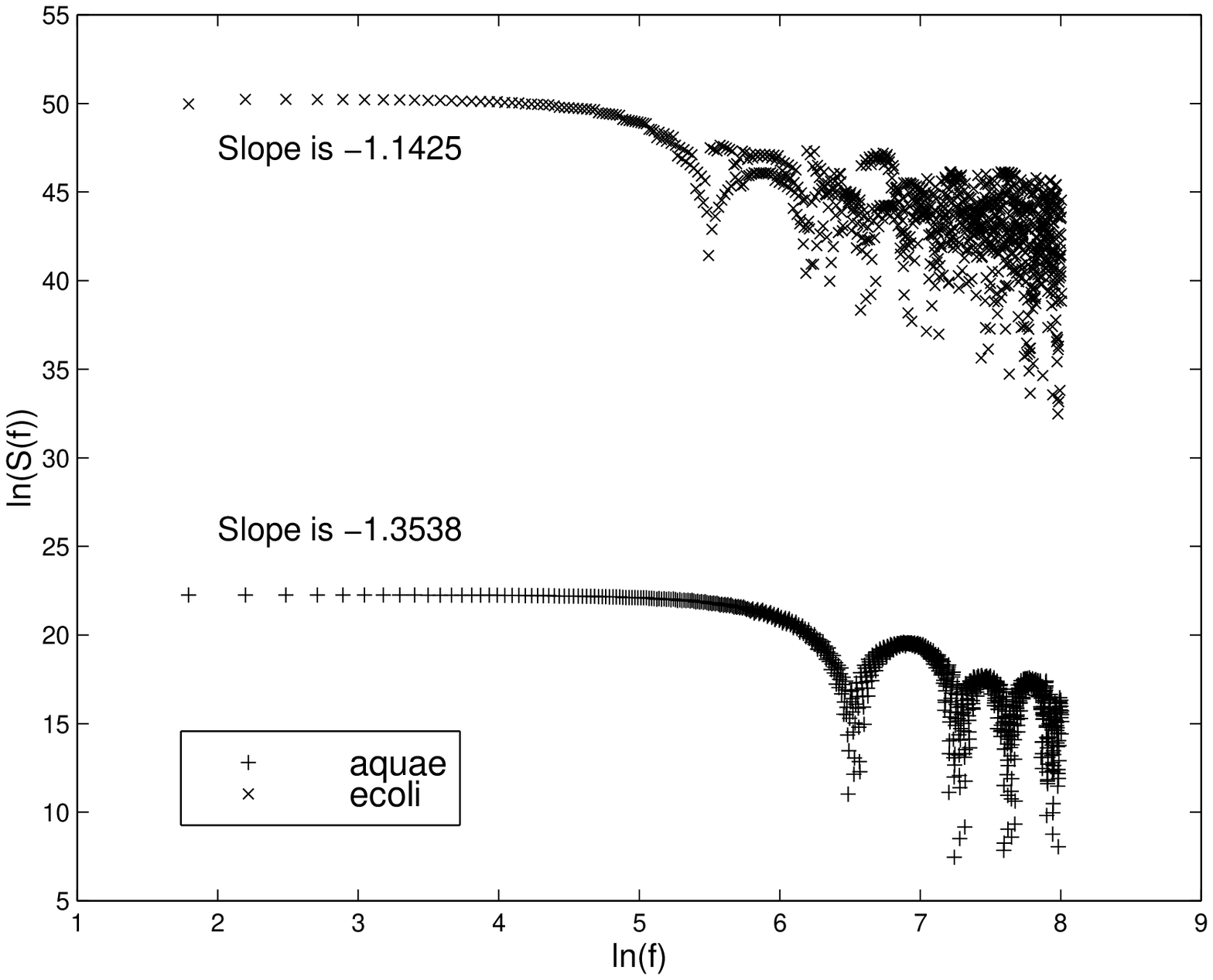}
\epsfxsize=8cm \epsfbox{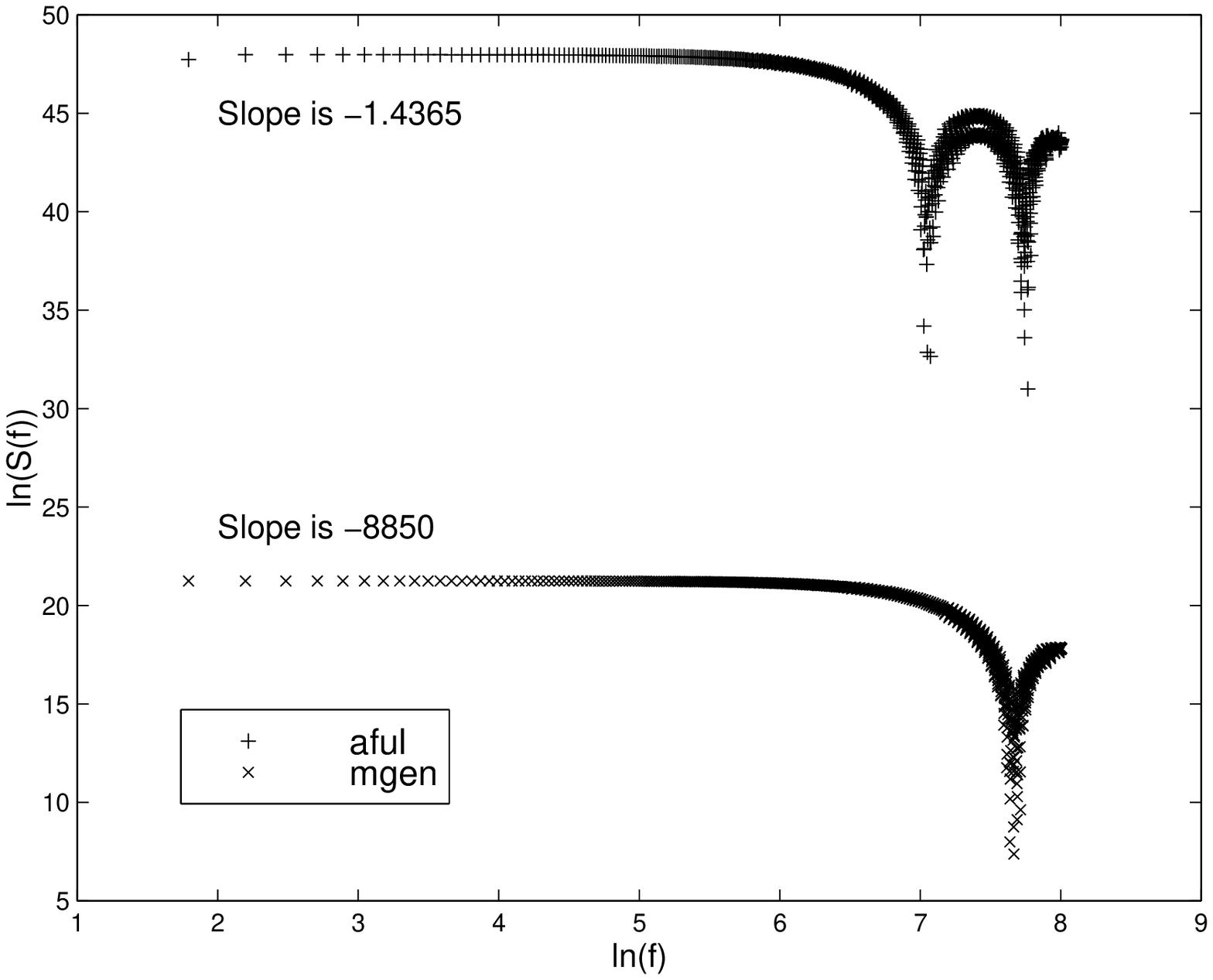}
  }
\caption{\footnotesize There exists $1/f$ noise in the interval of $f>1$. }
\label{noise}
\end{figure}

 From the right figure of FIG. \ref{M1pufig} it is seen that $S(f)$ does not
display clear power-law $1/f$ dependence on the frequencies when $f<1$.
Although the meaning of region $f>1$ of the power spectrum is not clear,
whether $S(f)$ displays perfect power-law $1/f$ in this region is important.
When one considers the electrical characteristics of polysilicon emitter
bipolar transistors, for high frequency analog applications the transistor
$1/f$ noise is also an important parameter since it can degrade the spectral
purity of the circuit$^{\cite{SDC96}}$. There is also some evidence that
$1/f$ noise spectral density in the low and in the high current region
have a different physical origion (the reader can refer Ref.\cite{SDC96}
and reference therein). 
 We
want to know if there is another region of frequencies in which $S(f)$
displays perfect power-law $1/f$ dependence on the frequencies. We carried
out the spectral analysis for $f>1$, and found that $S(f)$ displays almost a
perfect power-law $1/f$ dependence on the frequencies in some interval: 
\begin{equation}
S(f)\quad \propto \quad \frac{1}{f^{\beta }}.
\end{equation}
We give the results for coding length sequences of {\it M. genitalium, A.
fulgidus, A. aeolicus} and {\it E. coli} (their lengths are 303, 1538, 891 and 3034
respectively) in FIG. \ref{noise}, where we take $k$ values $%
f_{k}=3k\,\,(k=1,\cdots ,1000)$ of the frequency $f$. From FIG. \ref{noise}%
, it is seen that the length of the interval of frequency in which $S(f)$
displays almost a perfect power-law $1/f$ depends on the length of the
length sequence. The shorter sequence corresponds to the larger interval.
 
\begin{figure}
\centerline{\epsfxsize=5.8cm \epsfbox{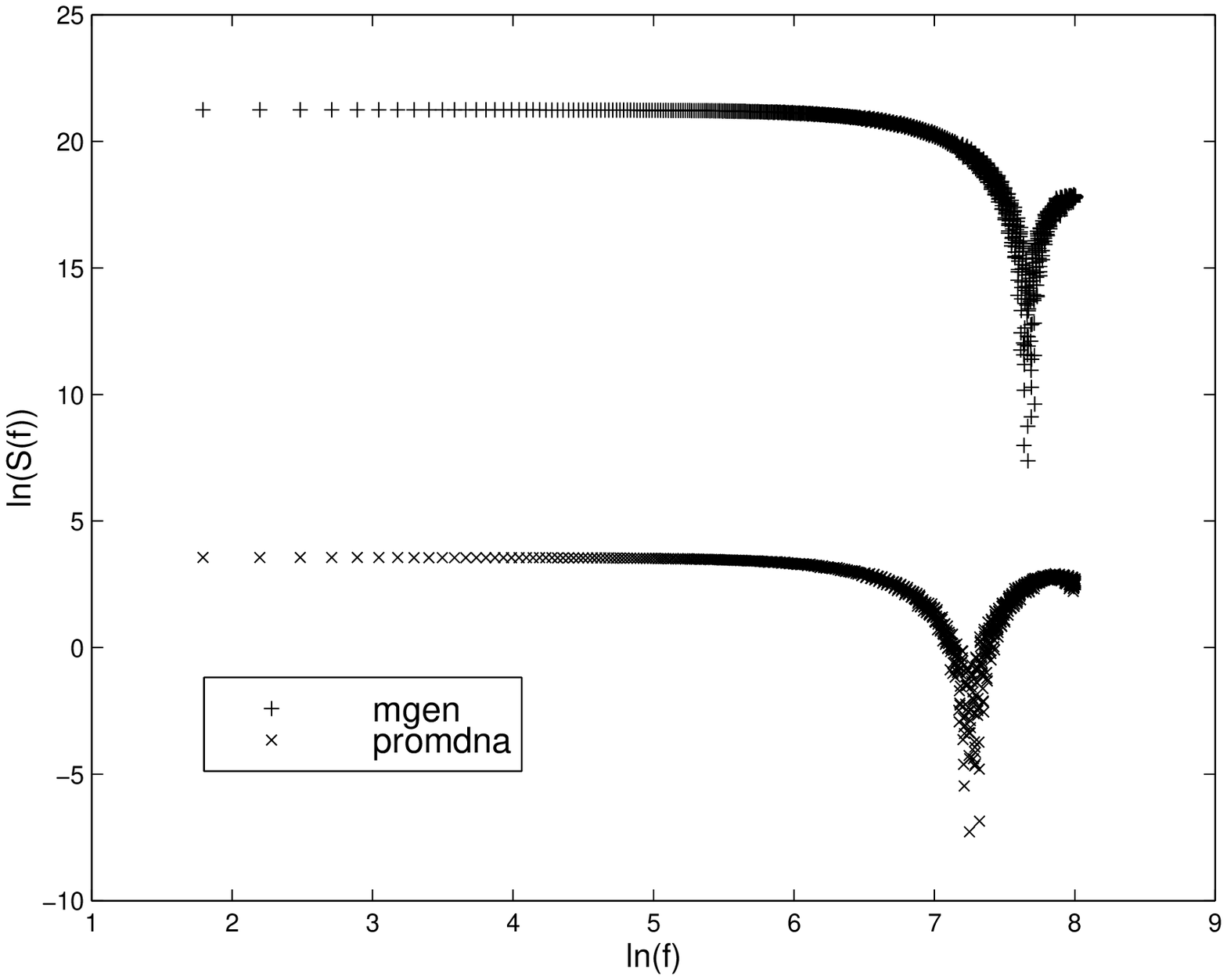}
\epsfxsize=5.8cm \epsfbox{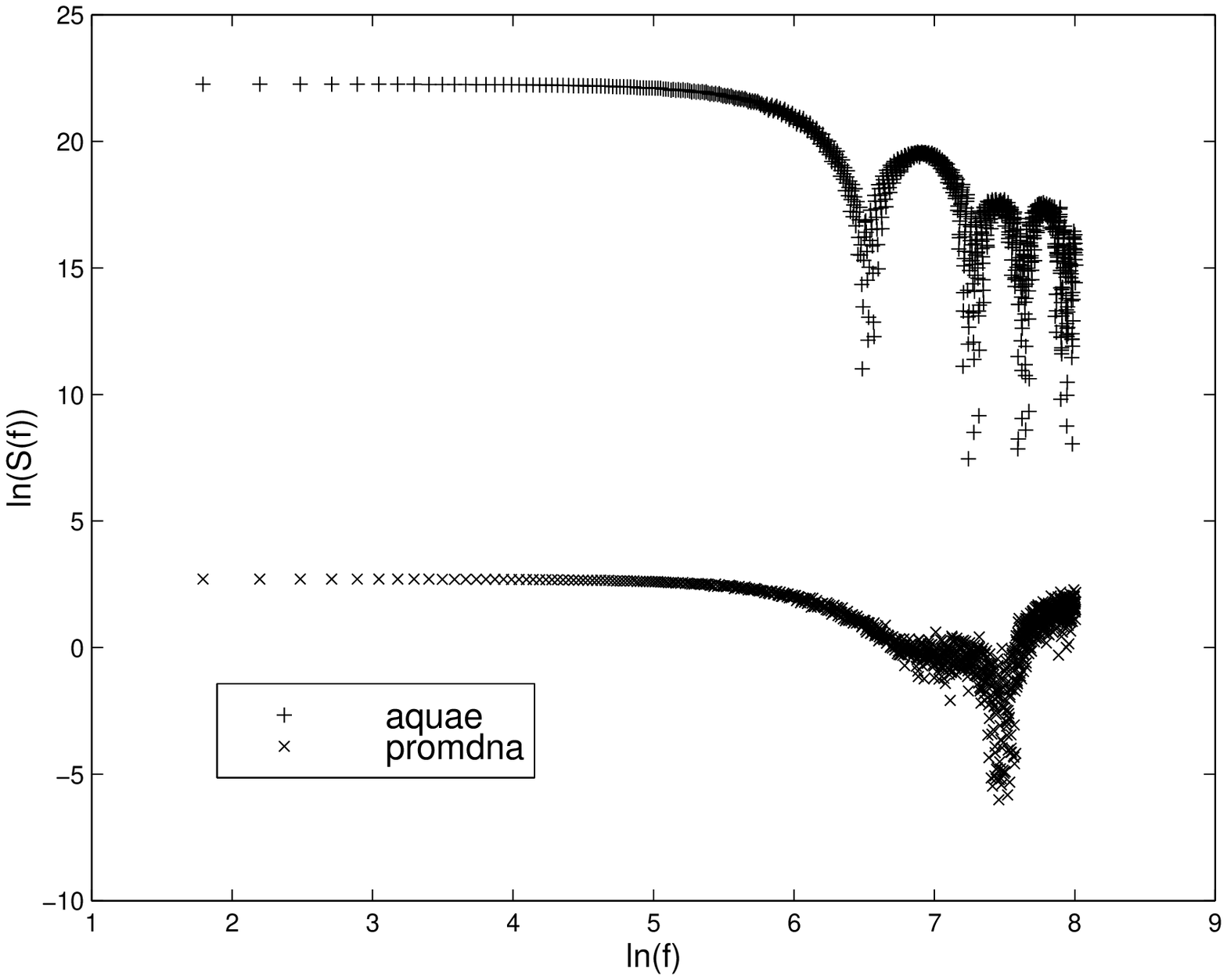}
\epsfxsize=5.8cm \epsfbox{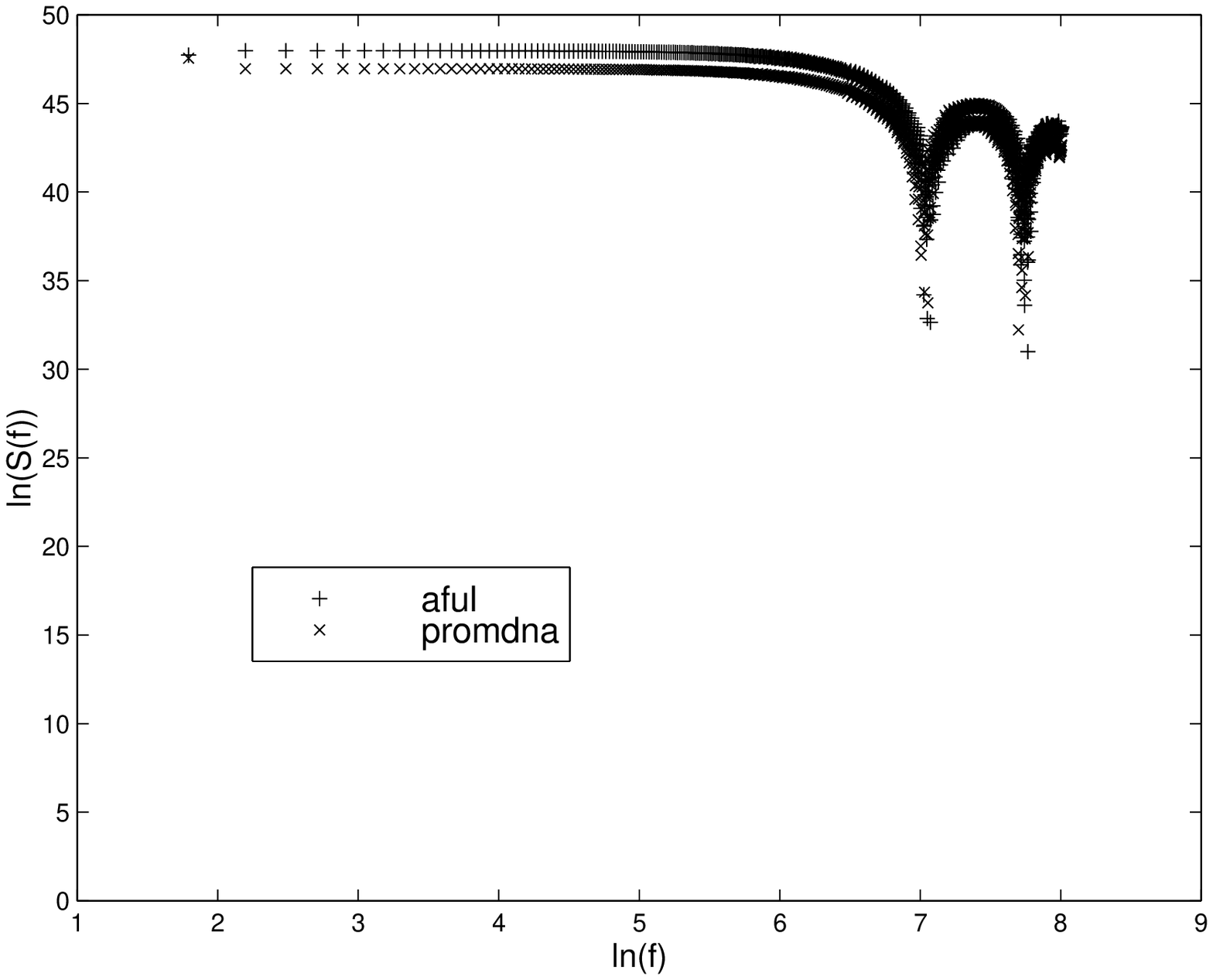}
  }
\caption{\footnotesize Compare the power spectral of length sequences and DNA sequences
when $f>1$. }
\label{comp}
\end{figure}

 From FIG. \ref{noise}, one can see that the power spectrum exhibit some
kind of periodicity. But the period seems to depend on the length of the
sequence. We also guess that the period also depends on the complexity of
the sequence. To support this conjecture, we got a promoter DNA sequence
from the gene bank, then replaced $A$ by -2, $C$ by -1, $G$ by 1 and $T$ by
2 (this map is given in \cite{YC}); so we obtained a sequence on alphabet $%
\{-2,-1,1,2\}$. Then a subsequences was obtained with the length the same as
the coding length sequences of {\it A. aeolicus, A. fulgidus} and {\it M. genitalium}
(their lengths are 891, 1538 and 303 respectively). A comparison is given in
FIG. \ref{comp}, but the results are not clear-cut.

\section{Discussion and conclusions}

\ \ \ \ Although the existence of the archaebacterial urkingdom has been
accepted by many biologists, the classification of bacteria is still a
matter of controversy$^{\cite{iwabe}}$. The evolutionary relationship of the
three primary kingdoms (i.e. archeabacteria, eubacteria and eukaryote) is
another crucial problem that remains unresolved$^{\cite{iwabe}}$.

 From Table \ref{jf}, we can roughly divide bacteria into two classes, one
class with $\alpha _{whole}$ less than 0.5, and the other with $\alpha
_{whole}$ greater than 0.5. All Archaebacteria belong to the same class
except {\it Pyrococcus abyssi}. All Proteobacteria belong to the same class except
{\it E. coli}; in particular, the closest Proteobacteria Helicobacter pylori 26695
and Helicobacter pylori J99 group with each other. In the class with $\alpha
_{whole}<0.5$, we have $\alpha _{cod}<\alpha _{noncod}$ except {\it H. pylori J99} and
{\it M. genitalium}; but in the other class we have $\alpha _{cod}>\alpha _{noncod}$.

 Using the exponent $\alpha^{\prime}$, we can also divide bacteria into two class
 as in Table \ref{ML}. In one class, $\alpha^{\prime}_{cod}<\alpha^{\prime} _{noncod}$.
 In another class, we have  $\alpha^{\prime}_{cod}>\alpha^{\prime} _{noncod}$ except
{\it Treponema pallidum} and {\it Borrelia burgdorferi}. Two
Hyperthermophilic bacteria {\it Aquifex aeolicus} and {\it Thermotoga 
maritima} group with each other.

  From Tables \ref{jf} and \ref{ML}, we can see the similar rules as above if
we use the exponents $\alpha_{cod}$ and $\alpha^{\prime}_{cod}$. This follows
the fact that the coding sequences occupy the main part of space of the DNA
chain of bacteria. This coincides with the conclusion of Ref.\cite{YA00}.

 Although from Table \ref{puM}, we can see the values of all $\beta$ are not
 far from $0$. From FIG.s \ref{jffig}, \ref{MLfig} and \ref{M1pufig}, one can
 see exponents $\alpha$ and $\alpha^{\prime}$ are more convincing than 
 the exponent $\beta $ because the error
of estimating $\alpha$ and $\alpha^{\prime}$ using the least-squares linear 
fit is much less than
that of the exponent $\beta$ (The values of $r$ in figures of obtaining exponents $\alpha$,
$\alpha^{\prime}$ and $\beta$ are 0.987685, 0.9949939 and 3.7918E-03 
respectively). From Tables \ref{jf} and \ref{ML}, we can see
most of values $\alpha$ and $\alpha^{\prime}$ are not equal to 0.5, hence we
can conclude that most of these length sequences exhibit long-range correlations.
We can also see the correlation have a rich variety of behaviours including
the presence of anti-correlations. Hence the length sequences have a same
character as the DNA sequences$^{\cite{Vie99}}$.
 Further more, from Table \ref{puM}, we get $\gamma
=1.0\pm 0.03$. Hence we can conclude that the long-range correlations
exist in most length sequences are linear. 

We find in an interval of frequency ($f>1$), $S(f)$ displays perfect
power-law $1/f$ dependence on the frequencies (see  FIG. 
\ref{noise}) 
\[
S(f)\quad \propto \quad \frac{1}{f^{\beta }}.
\]
The length of the interval of frequency in which $S(f)$ displays almost a
perfect power-law $1/f$ depends on the length of the length sequence. The
shorter sequence corresponds to the larger interval. The shape of the graph
of power spectrum in $f>1$ also exhibits some kind of periodicity. The
period seems to depend on the length and the complexity of the length
sequence.

\section*{ACKNOWLEDGEMENTS}
\ \ \   Authors Zu-Guo Yu and Bin Wang would like to express their thanks to 
Prof. Bai-lin Hao of Institute of Theoretical
Physics of Chinese Academy of Science for introducing them into this field and continuous
 encouragement. They also wants to thank Dr. Guo-Yi Chen of ITP  for useful
 discussions. Research is partially supported by Postdoctoral Research
 Support Grant No. 9900658 of QUT. The authors also want to thank the referee for
 telling the property of exponent $\gamma$ and many useful suggestions to improve
 this paper.

\onecolumn
\begin{table}
\caption{$\protect\alpha_{whole}$, $\protect\alpha_{cod}$ and $\protect\alpha%
_{noncod}$ of 21 bacteria.}
\label{jf}
\begin{center}
\begin{tabular}{|l|l|c|c|c|}
\ \ \ \ \ \ Bacteria & \ \ \ \ \ \ Category & $\alpha_{whole}$ & $%
\alpha_{cod}$ & $\alpha_{noncod}$ \\ \hline
Rhizobium sp. NGR234 & Proteobacteria & 0.24759 & 0.11158 & 0.34342 \\ 
Mycoplasma genitalium & Gram-positive Eubacteria & 0.37003 & 0.25374 & 
0.18111 \\ 
Chlamydia trachomatis & Chlamydia & 0.42251 & 0.37043 & 0.49373 \\ 
Thermotoga maritima & Hyperthermophilic bacteria & 0.43314 & 0.47659 & 
0.49279 \\ 
Mycoplasma pneumoniae & Gram-positive Eubacteria & 0.44304 & 0.45208 & 
0.49922 \\ 
Pyrococcus abyssi & Archaebacteria & 0.48568 & 0.39271 & 0.42884 \\ 
Helicobacter pylori J99 & Proteobacteria & 0.48770 & 0.43562 & 0.42089 \\ 
Helicobacter pylori 26695 & Proteobacteria & 0.49538 & 0.37608 & 0.41374 \\ 
Haemophilus influenzae & Proteobacteria & 0.49771 & 0.42432 & 0.53013 \\ 
Rickettsia prowazekii & Proteobacteria & 0.49950 & 0.33089 & 0.51923 \\ 
\hline\hline
Chlamydia pneumoniae & Chlamydia & 0.53982 & 0.53615 & 0.38085 \\ 
Methanococcus jannaschii & Archaebacteria & 0.54516 & 0.58380 & 0.34482 \\ 
M. tuberculosis & Gram-positive Eubacteria & 0.55621 & 0.57479 & 0.52949 \\ 
Aeropyrum pernix & Archaebacteria & 0.57817 & 0.63248 & 0.44829 \\ 
Bacillus subtilis & Gram-positive Eubacteria & 0.58047 & 0.59221 & 0.54480
\\ 
Borrelia burgdorferi & Spirochaete & 0.58258 & 0.53687 & 0.51815 \\ 
Archaeoglobus fulgidus & Archaebacteria & 0.59558 & 0.59025 & 0.46596 \\ 
Aquifex aeolicus & Hyperthermophilic bacteria & 0.59558 & 0.55964 & 0.43141
\\ 
Escherichia coli & Proteobacteria & 0.60469 & 0.62011 & 0.52000 \\ 
M. thermoautotrophicum & Archaebacteria & 0.62055 & 0.64567 & 0.38825 \\ 
Treponema pallidum & Spirochaete & 0.67964 & 0.70297 & 0.60914 \\ 
\end{tabular}
\end{center}
\end{table}

\begin{table}[tbp]
\caption{$\protect\alpha^{\prime}_{whole}$, $\protect\alpha^{\prime}_{cod}$ and 
$\protect\alpha^{\prime}_{noncod}$ of 21 bacteria.}
\label{ML}
\begin{center}
\begin{tabular}{|l|l|c|c|c|}
\ \ \ \ \ \ Bacteria & \ \ \ \ \ \ Category & $\alpha^{\prime}_{whole}$ & $%
\alpha^{\prime}_{cod}$ & $\alpha^{\prime}_{noncod}$ \\ \hline
Rhizobium sp. NGR234 & Proteobacteria & 0.17021 & 0.11223 & 0.28573 \\ 
Chlamydia trachomatis & Chlamydia & 0.172340 & 0.23801 & 0.66431 \\ 
M. tuberculosis & Gram-positive Eubacteria & 0.20185 & 0.18451 & 0.43716 \\ 
Mycoplasma genitalium & Gram-positive Eubacteria & 0.21632 & 0.25185 & 0.25971 \\ 
Escherichia coli & Proteobacteria & 0.25837 & 0.24567 & 0.62126 \\ 
Pyrococcus abyssi & Archaebacteria & 0.29809 & 0.18061 & 0.48169 \\ 
Bacillus subtilis & Gram-positive Eubacteria & 0.36791 & 0.46816 & 0.55325\\ 
Mycoplasma pneumoniae & Gram-positive Eubacteria & 0.37148 & 0.46475 & 0.46829 \\ 
Chlamydia pneumoniae & Chlamydia & 0.37216 & 0.26939 & 0.50734 \\ 
Rickettsia prowazekii & Proteobacteria & 0.41040 & 0.23109 & 0.50930 \\ 
Archaeoglobus fulgidus & Archaebacteria & 0.43149 & 0.35370 & 0.60835 \\ 
Helicobacter pylori 26695 & Proteobacteria & 0.44082 & 0.38500 & 0.39325 \\ 
\hline\hline
Haemophilus influenzae & Proteobacteria & 0.46121 & 0.44842 & 0.34842 \\ 
Aeropyrum pernix & Archaebacteria & 0.46203 & 0.45520 & 0.24850 \\ 
M. thermoautotrophicum & Archaebacteria & 0.48038 & 0.48870 & 0.36249 \\ 
Thermotoga maritima & Hyperthermophilic bacteria & 0.49453 & 0.50457 & 0.27005 \\ 
Aquifex aeolicus & Hyperthermophilic bacteria & 0.50237 & 0.50582 & 0.31488\\ 
Helicobacter pylori J99 & Proteobacteria & 0.54547 & 0.50999 & 0.48640 \\ 
Treponema pallidum & Spirochaete & 0.56357 & 0.56808 & 0.65350 \\
Borrelia burgdorferi & Spirochaete & 0.61186 & 0.58016 & 0.61772 \\ 
Methanococcus jannaschii & Archaebacteria & 0.72726 & 0.73384 & 0.33780 \\ 
\end{tabular}
\end{center}
\end{table}

\begin{table}[tbp]
\caption{$\protect\beta_{whole}$, $\protect\beta_{cod}$ and $\protect\beta%
_{noncod}$; $\protect\gamma_{whole}$, $\protect\gamma_{cod}$ and $\protect%
\gamma_{noncod}$ of 21 bacteria.}
\label{puM}
\begin{center}
\begin{tabular}{||l||c|c|c||c|c|c||}
\ \ \ \ \ \ Bacteria & $\beta_{whole}$ & $\beta_{cod}$ & $\beta_{noncod}$ & $%
\gamma_{whole}$ & $\gamma_{cod}$ & $\gamma_{noncod}$ \\ \hline
M. genitalium & 0.05880 & 0.02030 & -0.00708 & 1.00017 & 0.99698 & 1.01652
\\ \hline
H. pylori 26695 & 0.05026 & -0.01412 & 0.01196 & 0.99902 & 1.00057 & 0.99538
\\ \hline
M. jannaschii & 0.04850 & -0.02640 & -0.12547 & 0.99727 & 0.99079 & 0.99767
\\ \hline
C. pneumoniae & 0.04405 & 0.01071 & -0.01906 & 0.99998 & 1.00099 & 0.99348
\\ \hline
A. aeolicus & 0.03152 & 0.00811 & -0.00115 & 1.00441 & 0.99816 & 0.99870 \\ 
\hline
H. pylori J99 & 0.01968 & 0.04512 & -0.05815 & 0.99867 & 0.99926 & 0.99349
\\ \hline
T. maritima & 0.00737 & -0.02656 & 0.01965 & 0.99726 & 0.99524 & 0.98866 \\ 
\hline
C. trachomatis & 0.00256 & -0.05829 & -0.02549 & 0.99767 & 1.00211 & 0.98553
\\ \hline
R. sp. NGR234 & 0.00230 & 0.04048 & -0.10905 & 1.00570 & 0.99612 & 1.01515
\\ \hline
M. thermoauto. & -0.00217 & -0.11916 & 0.02079 & 1.00479 & 1.00171 & 1.00063
\\ \hline
T. pallidum & -0.00422 & -0.02902 & 0.09510 & 1.01009 & 1.01532 & 1.00222 \\ 
\hline
M. pneumoniae & -0.01137 & 0.03437 & -0.05573 & 0.98820 & 0.98783 & 0.97260
\\ \hline
P. abyssi & -0.01589 & -0.04242 & 0.00071 & 0.99888 & 0.99816 & 0.99293 \\ 
\hline
E. coli & -0.01917 & -0.05513 & 0.01772 & 0.99856 & 1.00197 & 0.98938 \\ 
\hline
M. tuberculosis & -0.02653 & -0.05653 & -0.02698 & 1.00062 & 0.99974 & 
1.00801 \\ \hline
A. pernix & -0.03882 & 0.01648 & -0.09395 & 1.00298 & 1.00407 & 1.00286 \\ 
\hline
B. burgdorferi & -0.04420 & -0.05189 & -0.10710 & 0.99287 & 0.99792 & 1.03206
\\ \hline
R. prowazekii & -0.04884 & -0.12438 & -0.07581 & 1.00284 & 0.99043 & 0.99991
\\ \hline
H. influenzae & -0.05338 & -0.04853 & -0.04341 & 0.99798 & 1.00248 & 0.98684
\\ \hline
A. fulgidus & -0.06372 & -0.08130 & -0.00881 & 1.00347 & 1.00610 & 0.98219
\\ \hline
B. subtilis & -0.06887 & -0.17231 & -0.02380 & 0.99629 & 1.00853 & 0.98666
\\ 
\end{tabular}
\end{center}
\end{table}

\end{document}